\documentclass[sigconf, nonacm]{acmart}
\usepackage{todonotes}
\usepackage{multirow}
\usepackage{algorithm2e}
\newtheorem{theorem}{Theorem}
\newtheorem{definition}{Definition}

\AtBeginDocument{%
  \providecommand\BibTeX{{%
    \normalfont B\kern-0.5em{\scshape i\kern-0.25em b}\kern-0.8em\TeX}}}

\begin{document}

\title{LANNS: A Web-Scale Approximate Nearest Neighbor Lookup System}

\author{Ishita Doshi}
\affiliation{%
  \institution{Linkedin}
  \city{Bangalore}
  \country{India}
}

\author{Dhritiman Das}
\affiliation{%
  \institution{Linkedin}
  \city{Bangalore}
  \country{India}
}

\author{Ashish Bhutani}
\affiliation{%
 \institution{Linkedin}
 \city{Bangalore}
 \country{India}
}

\author{Rajeev Kumar}
\affiliation{%
  \institution{Linkedin}
  \city{Bangalore}
  \country{India}
}

\author{Rushi Bhatt}
\affiliation{%
  \institution{Linkedin}
  \city{Bangalore}
  \country{India}
}

\author{Niranjan Balasubramanian}
\affiliation{%
  \institution{Linkedin}
  \city{Bangalore}
  \country{India}
}

\renewcommand{\shortauthors}{Doshi et al.}

\begin{abstract}
 Nearest neighbor search (NNS) has a wide range of applications in information retrieval, computer vision, machine learning, databases, and other areas. Existing state-of-the-art algorithm for nearest neighbor search, Hierarchical Navigable Small World Networks (HNSW), is unable to scale to large datasets of 100M records in high dimensions. In this paper, we propose LANNS, an end-to-end platform for Approximate Nearest Neighbor Search, which scales for web-scale datasets. Library for Large Scale Approximate Nearest Neighbor Search (LANNS) is deployed in multiple production systems for identifying top-K (100 $\leq$ k $\leq$ 200) approximate nearest neighbors with a latency of a few milliseconds per query, high throughput of ~2.5k Queries Per Second (QPS) on a single node, on large (~180M data points) high dimensional (50-2048 dimensional) datasets.
\end{abstract}




\maketitle

\section{Introduction}
Nearest-neighbor search is an effective technique for information retrieval and several other machine learning applications. Despite its simplicity and wide-ranging utility, efficiently building and serving k-nearest neighbor data structures to web-scale has remained a challenge. In this paper, we describe a system called LANNS (Large Scale Approximate Nearest Neighbor Search), designed and deployed in a web-scale environment at LinkedIn. The LANNS system has been deployed in a production environment for identifying top-K (with k ranging from 100-200) approximate nearest neighbors with very low latency (few milliseconds per query), very high throughput (roughly ~2.5K Queries Per Second (QPS) on a single node), on large (e.g., 180M data points) high dimensional (e.g., 128, 256, or 2048 dimensional) data sets.  

\begin{figure}[tb]
    \centering
    \includegraphics[width=\linewidth]{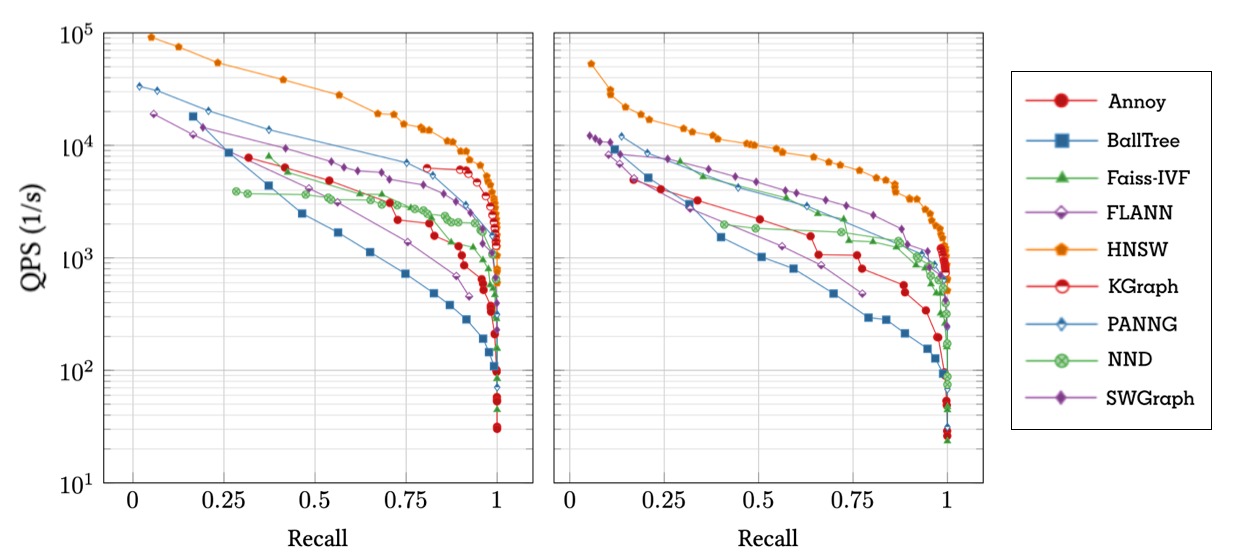}
    \caption{Recall v/s QPS on SIFT1M. Left: 10 nearest neighbors, Right: 100 nearest neighbors}
    \label{fig:annBenchmark}
\end{figure}

Broadly, nearest neighbor search approaches fall into four categories. They can be tree based\cite{kd, rpTrees, annoy, apdTrees}, product quantization based\cite{PQ, ge2013optimized, heo2014distance, kalantidis2014locally}, {Locality Sensitive Hashing (LSH) based\cite{lsh, andoni2018data, bai2014data, DBLP:journals/corr/AndoniR15, weiss2009spectral}}, or graph based\cite{HNSW, swGraph, panng, kgraph}. Most of the highly scalable methods return approximate nearest neighbors (i.e., miss out on some of the k-nearest neighbors in the results) in order to speed up the search. The recall, measured as the fraction of true $k$-nearest neighbors returned in a result set of size $k$, is generally traded off for the query latency or throughput.  Figure~\ref{fig:annBenchmark}\cite{annbenchmarks}, shows such a compromise between various state of the art algorithms (Annoy\cite{annoy}, BallTree\cite{ballTree}, Faiss-IVF\cite{faiss, PQ}, FLANN\cite{flann}, Hierarchical Navigable Small World (HNSW) graph\cite{HNSW}, KGraph\cite{kgraph}, PANNG\cite{panng}, PyNNDescent\cite{nnd} and SWGraph~\cite{swGraph}) on the SIFT1M dataset. It is evident from Figure~\ref{fig:annBenchmark}, and other offline benchmarks conducted by us, that HNSW tends to outperform its competitors considering QPS vs recall tradeoff. 

We have used HNSW as the core approximate nearest neighbor (ANN) algorithm inside LANNS. However, LANNS has been built to be extensible to support other ANN algorithms in the future {with a bounded drop in recall}.

Despite the favorable performance characteristics and popularity of HNSW\cite{HNSW}, building the HNSW data structure does not scale well for large, high dimensional datasets. For example, building the HNSW index on a real dataset of size $2.7M$ with $256$ dimensions takes about \textbf{2 hours 20 minutes} on a single machine. At LinkedIn, we often have to serve $k$-NN queries on datasets containing \textbf{100M-500M} records with dimensionality of \textbf{128-2048}. This renders the default single-machine HNSW index build methods impractical. Furthermore, beyond a certain index size, procurement and maintenance cost of high memory servers also increases compared to commodity hardware. It is therefore necessary to be able to split up the dataset into multiple shards.

{In this paper, we present LANNS, our end-to-end platform currently in production at LinkedIn, which enables web-scale nearest neighbor search in a variety of applications. As part of LANNS, we propose a two-level data partitioning strategy that allows us to scale the HNSW algorithm to web-scale datasets at index build time, as well as for online serving. {We show that using this parallel building of separate HNSW indices, one for each data partition, and flexible data segmentation, we achieve fast index build and online serving. Our proposed data segmentation techniques also bound the drop in recall as compared to the HNSW algorithm.} These segmentation techniques have theoretical guarantees on the recall as a function of the tuneable partitioning parameters. These guarantees are on similar lines as ~\cite{rpTrees}. We demonstrate the empirical performance of our proposed strategy on two open-source and four real-world datasets.}

\subsection{Our Contributions}
Our contributions in this paper are as follows:
\begin{enumerate}
    \item We propose a two-level data partitioning strategy that allows us to scale HNSW indexing to web-scale datasets.
    \item We propose a flexible data segmentation framework within each partition which allows further scaling. We propose two segmentation strategies with guarantees for a bounded drop in recall as a function of data size. 
    \item We show, through extensive benchmarking, that for a majority of queries, high recall is achieved while querying only one or a few segments. In other words, we show that our partitioning and segmentation framework performs and scales well.
    \item We demonstrate the performance of our end-to-end system on various open-source and real datasets and show its favorable scalability properties.
\end{enumerate}

The rest of the paper is organized as follows. We discuss the related work in Section~\ref{sec:rw}, followed by a brief overview of the HNSW algorithm in Section~\ref{sec:hnsw}. In Section~\ref{sec:scaling}, we motivate our two-level data partitioning, followed by the partitioning strategies. In Section~\ref{sec:framework}, we describe the Spark framework, and also briefly describe scaling Brute Force Search for web-scale datasets in Section~\ref{sec:bfs}. In Section~\ref{sec:expt}, we present our experimental results on open source as well as real datasets, followed by a brief discussion on our online framework in Section~\ref{sec:online}. We conclude and discuss future work in Section~\ref{sec:conclusion}.

\section{Related Work}
\label{sec:rw}
In this section, we will discuss the techniques and algorithms used for nearest neighbor search, as well as some works similar to LANNS. 

\textbf{Annoy}\cite{annoy} -- A tree-based method for approximate nearest neighbor search algorithm proposed by Spotify for their music search engine. They build forests of trees by recursively splitting the dataset. Each tree is constructed by picking two points at random and splitting the dataset using the hyperplane separating the two points. This is done recursively until the number of points in space is small enough to perform an exhaustive search. This method has several advantages-- (i) Parameters can be tuned to change the accuracy v/s speed trade-off, (ii) Index can be easily serialized and stored as static files. This enables the sharing of index across processes. However, these algorithms give low recall when the queries are near the boundaries or the splitting hyperplanes.

\textbf{Locality Sensitive Hashing (LSH)}\cite{lsh} -- LSH is a hashing based technique where points are assigned to buckets such that, with high probability, similar points are found within the same bucket, while points far from each other are likely to be in different buckets. Variants of LSH can be data dependent\cite{andoni2018data, bai2014data, DBLP:journals/corr/AndoniR15, weiss2009spectral} or data independent. This method builds the index in linear time and has good theoretical guarantees of sub-linear query time, however, for adversarial data, this algorithm might run as slow as a linear scan. In related work, Dasgupta\cite{rpTrees} proposed a spatial data structure called Random Projection Tree. This technique builds a tree by recursively partitioning the data using random hyperplanes. \cite{rpTrees} also talks about the case where queries are near the splitting hyperplanes. They propose to use physical or virtual spills, i.e., route data points to multiple partitions based on their distance to the splitting hyperplane, or, route queries to multiple partitions based on the distance to the splitting hyperplane. In \cite{apdTrees}, the authors propose Approximate Principal Direction Trees, another spatial data structure that also builds a tree by recursively splitting the data points using a hyperplane. They propose to use an approximate eigenvector to split the dataset and claim that their proposed method reduces the average diameter at the same rate as PCA Trees~\cite{pca}, and has a lower runtime. 

\textbf{Product Quantization (PQ)}\cite{PQ, ge2013optimized, heo2014distance, kalantidis2014locally} -- PQ is a compression-based approximate nearest neighbor search method. The main motivation behind PQ is to compress the space into a product of lower dimension spaces and to quantize each of these subspaces separately. The dataset is split into multiple smaller, tall datasets based on its dimensions, and each of these sub-datasets are then clustered into $k$ clusters. Since all data points are present in each of the sub-datasets, we receive multiple codes, and one data point is represented by a list of these cluster codes. One advantage of this method is the compression of datasets and this results in significant speedup. However, in this approach as well, the exact nearest neighbors might lie in other clusters.

\textbf{Sparsest Cut and Eigenvectors}\cite{trevisan2013lecture} -- Sparsest cut aims to partition the vertices of a graph in a way that the weights of the edges cut during this partitioning are as small as possible. This is typically done by using the Laplacian of the adjacency matrix of the graph and using the second smallest eigenvector of the same. \cite{trevisan2013lecture} shows that using the second smallest eigenvector of the Laplacian has some proven theoretical guarantees.

\textbf{Hierarchical Navigable Small World(HNSW)}\cite{HNSW} -- A graph-based technique built on the idea of Small Worlds (SW). Suppose you build a hierarchy of SW graphs that separate links according to their lengths. At earlier stages of the search, you traverse long edges and zoom into a local minima for the query, and at later stages, you search the neighborhood of the local minima to find the nearest neighbors to the query. This method has the benefit of tuning parameters to adjust the accuracy v/s speed trade-off, and the space v/s speed trade-off. It has a polylogarithmic time complexity and is highly competitive on real-world datasets\cite{annbenchmarks}. However, the HNSW indexing is not scalable for large datasets. We extend this work to scale to large datasets with an implementation in Apache Spark\cite{spark} and employing various techniques motivated by Random Projection Trees\cite{rpTrees}, Approximate Principal Direction Trees\cite{apdTrees} and Sparsest Cuts\cite{trevisan2013lecture}.

\section{Hierarchical Navigable Small Worlds (HNSW)}
\label{sec:hnsw}
\begin{figure}[ht]
    \centering
    \includegraphics[width=0.8\linewidth]{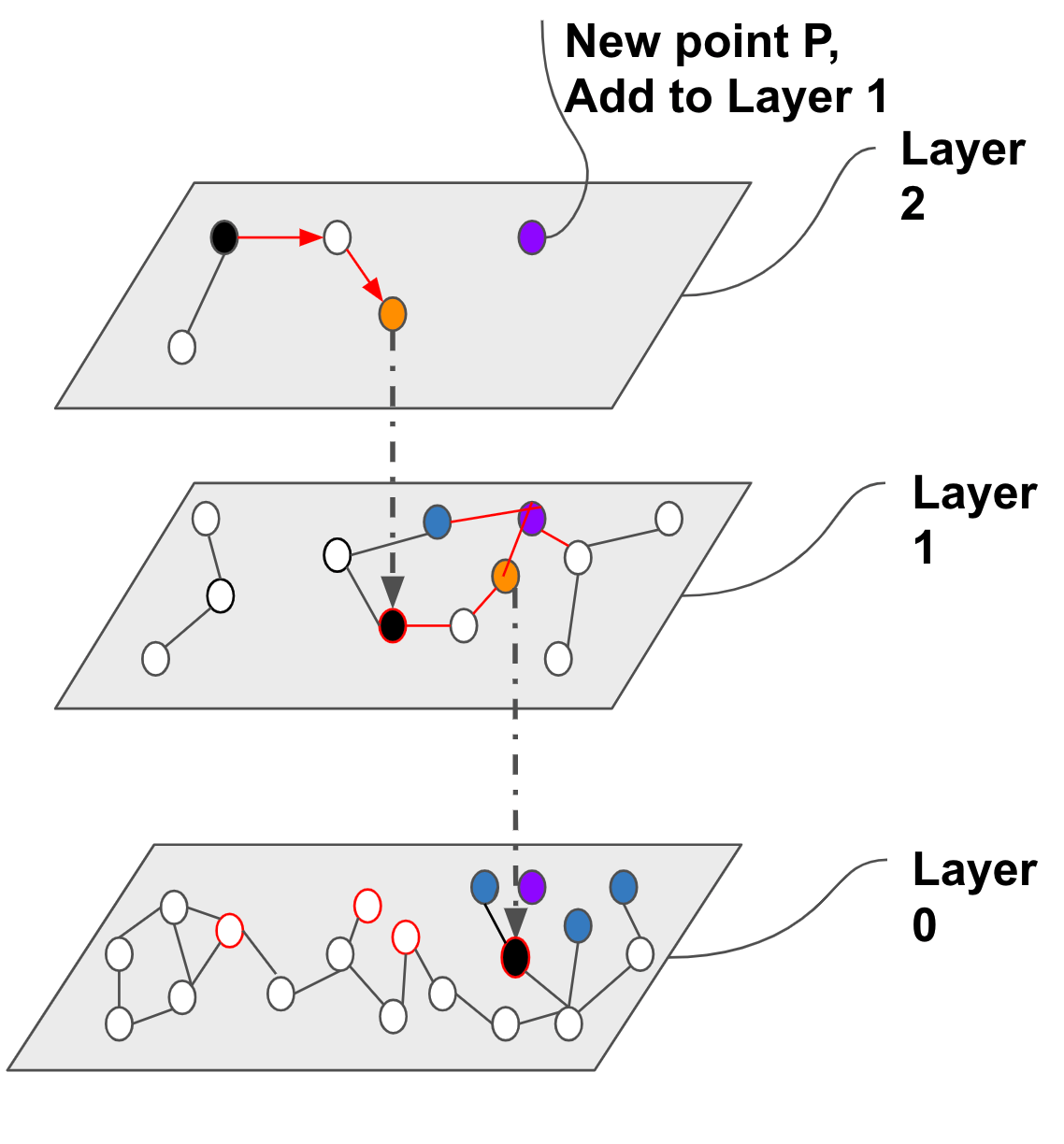}
    \caption{Ingestion in HNSW}
    \label{fig:hnsw}
\end{figure}

The key idea behind HNSW algorithm is to build a hierarchy of graphs that separate links according to their lengths. HNSW is a multi-layered and multi-resolution variant of a proximity graph. In all layers, the average connections per element can be limited. The insertion process is depicted in Figure~\ref{fig:hnsw}. The base layer, layer zero, consists of all elements. Each higher-level has a smaller number of points. For each data point, a maximum level, $m$, is chosen at random using a power distribution (Layer 1 in Figure~\ref{fig:hnsw}). The point is added to all layers from $m$, down to $0$. The first part starts at the topmost level down to level $m+1$ by greedily traversing the graph to find the closest neighbor in the layer. This closest neighbor is used as the entry point to the next layer. The second part starts at level m down to level 0. At each layer, $M_{layer}$ number of nearest points are found and connected to the point to be added. 

For the searching phase, from the topmost to the base layer, we find the closest point in the layer. For each of these layers, we find the closest point, use it as the entry point to the next layer. {For layer $0$, i.e., the base layer, we obtain $c\: (c \geq N)$ candidates, where $N$ is the number of nearest neighbors to be found. These $c$ candidates are then filtered to return the best $N$ candidates.}

\section{Two-Level Partitioning}
\label{sec:scaling}

In many real use cases, we often require the algorithm to scale to datasets of size 100M-500M. The state-of-the-art, HNSW algorithm, takes about 2 hours 20 minutes to index a dataset of 2.7M records, becomes infeasible in the real-world scenarios. {Often these indices are used in production systems that do not have the capacity to support such large datasets. }

We propose a horizontal, two-level partitioning of the data such that each partition represents a subset of the dataset. We attain acceptable indexing times by building a separate HNSW index within each partition, and we host one, or a few, partitions on each online server node. The two-levels of partitioning, \texttt{sharding} and \texttt{segmentation}, are two different dimensions that help in solving two aspects of the scaling problem.

\subsection{Sharding}

Sharding, our first level of partitioning, is necessary for a very large dataset where the memory requirements for keeping the entire dataset is large and cannot be accommodated in a single node. For example, consider a dataset with 50M records in a 500-dimensional space. The data storage would be approximately \textbf{93G}. In addition to this 93G required for storing the data, we also need to consider the memory requirements of building a graph. Assuming that the total storage required would be about 128G, an index of this size would not fit in a production node with a standard memory configuration of 64G. Building customized production nodes with higher memory are not feasible since the cost per GB increases super-linearly with total machine memory due to the higher cost of compatible components, higher failure rates, etc. Thus, we propose our first level of partitioning as \texttt{sharding}. {When a point is inserted, it is hashed to \textbf{one} particular shard using the key of the data point. Since this partitioning does not exploit any locality information, each query is routed to \textbf{all} shards of the LANNS index. The response from each shard is aggregated by merging all shard level candidates and picking the topK best candidates.}

\texttt{Sharding} allows us to scale horizontally by partitioning the dataset. Each shard is hosted on a separate server node, which in turn enables us to keep the memory requirement of a single server node under control. This also allows us to use standard configuration server nodes with 64G memory.

Let us consider a scenario where the server node has enough memory but the indexing time is unacceptable for the use case. In such cases, with only one level of partitioning, we would create more shards. There is an additional merge cost involved at the master/broker or the system which makes calls to the shards. Higher the number of shards, higher is the merge cost. The master, which could be a system with low memory of 2G-4G, would need to merge the results from these shards and give the final topK responses. Considering the build time, one may use a large number of shards which could mean that an increased merge cost, and possibly require higher memory for use cases with a large number of shards. Having a large number of shards also comes with an additional undesirable operational cost of maintaining a large number of systems in production, and increased hardware footprint in terms of the cluster (collection of server nodes) size. 

\subsection{Segmentation}
Segmentation, our second level of partitioning aims at reducing the disadvantages of sharding. Each shard is further split into smaller partitions called segments. This segmentation can be done using the same techniques like sharding, or smarter segmentation techniques which allow a query to be routed to one or only a few segments. Routing to a single segment during query retrieval may have a negative impact on the topK recall, but smarter segmentation strategies can be employed to keep this impact bounded. These smarter strategies are typically learnt using the data. We describe two of our "smarter" segmentation strategies in Section~\ref{sec:hyperplanes}. {Employing these same techniques in sharding becomes complicated as the online service employs an external broker in front of the shards, which are not co-located. }

{Another added advantage of segmentation is that segmentation provides the same scalability as sharding for offline ingestion. This is particularly useful for cases where the dataset is small enough to fit into a single server node, but it is large enough to render the HNSW indexing time unacceptable.} This helps in avoiding setting up a multi-sharded setup till the time the dataset becomes large enough. For a large dataset, this also enables us to keep the number of shards under control. It is worthwhile to note that indexing is done offline for online serving as well. Each partition, i.e., each segment is built separately and in parallel. Thus, segmentation does not hamper the scalability of indexing. Since multiple segments are hosted within the same server node, it also reduces the online hardware footprint. 

As mentioned earlier, routing to one or a few segments can cause a drop in the recall. We propose segmentation techniques that bounds this drop in the recall. Another point to note is that in cases where a query is routed to multiple segments, there is an additional merge cost. For our online serving systems, this merge happens within the shard and does not require additional network I/O to send results from each shard to the broker node.

\subsection{Segmentation Strategies}
\label{sec:hyperplanes}
In this section, we describe three types of segmenters, the Random Segmenter (RS), Random Hyperplane Segmenter (RH), and Approximate Principal Direction Hyperplane Segmenters (APD). The random segmenter is a data-independent segmenter, the random hyperplane and approximate principal direction hyperplane segmenters are data-dependent segmenters.

\subsubsection{Random Segmenter (RS)}
In this particular segmenter, no type of learning from data is required. The segmenter is essentially a modulo segmenter. At indexing time, for each document, it randomly selects a segment where it should be routed. Since this type of segmenter has no guarantees about the locality of the data, a query vector would be routed to \textbf{all} segments. 

\subsubsection{Random Hyperplane Segmenters(RH)}
\label{sec:RH}

Random Hyperplane Segmenters, motivated by Randomized Projection Trees\cite{rpTrees}, builds a short tree of hyperplanes. The motivation behind this work is the following-- If two points are similar, they would be close in the space, and it is highly unlikely that a randomly chosen hyperplane would split the two. However, if two points are far, there would be a high probability that the two points would be split. This enforces a sense of locality. With high probability, points similar to each other would lie in the same partition. We exploit this intuition to design our segmenters as-- at each internal node of our segmenter, we first generate a random hyperplane from the unit sphere and project all points on this generated hyperplane. We then perform a median split based on these projected values. 

However, with a low probability, this method faces the problem of missing nearest neighbors that lie across the boundary in the other partition. We employ the method of ``virtual`` spill, where we maintain a left and right boundary around the splitting point. When a query point arrives and it lies within these left and right boundaries, we route the queries to both partitions. \footnote{Note that instead of using a virtual spill, we can also perform data side spill during ingestion, where data points lying within the left and right boundaries are routed to both partitions.}

\begin{figure}[ht]
    \centering
    \includegraphics[width=0.5\linewidth]{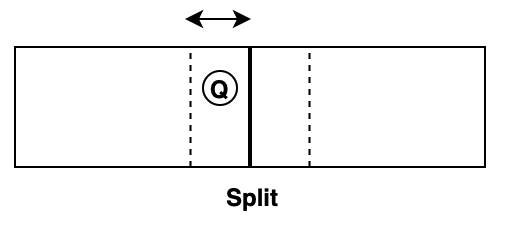}
    \caption{Virtual Spill where Q is the query point lying close to the splitting hyperplane}
    \label{fig:splitting}
\end{figure}

We briefly describe the insertion and querying process and then state the theoretical bounds provided in \cite{rpTrees}. 
Let the data set be represent by a matrix $\mathcal{D}$ of size $n \times d$, where $n$ is the number of points and $d$ is the number of dimensions, and $\alpha$ be the amount of spill. Let $x.h$ refer to the projection of $x$ on $h$, and $U$ denote the $n$ dimensional vector of projections, $U = \mathcal{D}.h$. For insertion of a point $x$, if $x.h < median(U)$ route to the left partition, else route to the right. For query of a point $q$, let $l = 0.5 - \alpha$ fractile point in $U$, and $r = 0.5 + \alpha$ fractile point in $U$. If $q.h < l$, route to the left, if $q.h > r$, route to the right, else route to both sides. 

We state results from \cite{rpTrees} which are directly applicable to our RH segmenter.

\begin{definition}
For a query, $q$, and data points $x_1, \hdots x_n$, and let $x_{(1)}, \hdots, x_{(n)}$ denote the reordering of points by increasing distance from $q$. Let us consider the potential function for $1$-nearest neighbor.

\begin{equation}
    \Phi_m(q, {x_1, \hdots x_n}) = \frac{1}{m}\sum_{i=2}^{n} \frac{||q - x_{(1)}||}{||q-x_{(i)}||}
\end{equation}

Generalizing the potential function for $k$-nearest neighbors:

\begin{equation}
    \Phi_{k,m}(q, {x_1, \hdots x_n}) = \frac{1}{m}\sum_{i=k+1}^{n} \frac{\sum_{j=1}^{k} ||q - x_{(j)}|| / k}{||q-x_{(i)}||}
\end{equation}

\end{definition}

\begin{theorem}
\label{thm:randomProj}
Suppose we build a tree on data points $x_1, \hdots x_n$ of depth $\mathcal{L}$, with $\alpha$ spill. If this tree is used to find the nearest neighbors of a query $q$, then the probability that it fails to return $x_{(1)}$ is at most:
\begin{equation}
    \frac{1}{2\alpha}\sum_{i=0}^{\mathcal{L}} \Phi_{(0.5+\alpha)^in}(q, {x_1, \hdots, x_n})
\end{equation}
and it fails to return $x_{(1)}, \hdots, x_{(k)}$ is
\begin{equation}
    \frac{k}{\alpha}\sum_{i=0}^{\mathcal{L}} \Phi_{(k, (0.5+\alpha)^in)}(q, {x_1, \hdots, x_n})
\end{equation}
\end{theorem}

\begin{figure}[ht]
    \centering
    \includegraphics[width=0.9\linewidth]{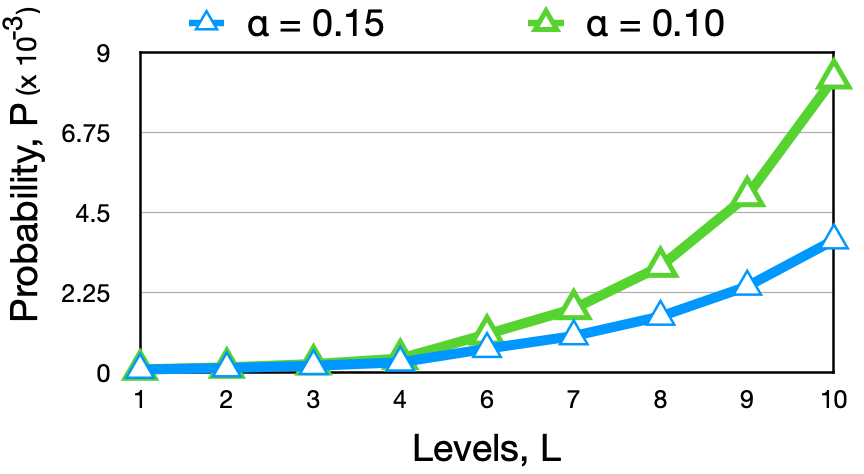}
    \caption{Probability of Failure with increasing Levels}
    \label{fig:probOfFailure}
\end{figure}

As we increase the depth of the tree, the number of hyperplanes used also increases. As more hyperplanes are used, there is a higher probability of two close points being separated by the segmentation algorithm. In Figure~\ref{fig:probOfFailure}, we approximate the probability that our segmentation algorithm fails to return $x_(1)$. We assume that $\Phi'_{m}(q, {x_1, \hdots, x_n}) \approx \frac{1}{2\alpha}$ since the second part of the calculation is data-dependent, and $P(\mathcal{L}) \approx \sum_{L=1}^{\mathcal{L}} \frac{1}{2(0.5+\alpha)^in}$. Also, for the ease of demonstration, let $n=10000$. As we use more levels, there is a higher probability of missing out on the exact nearest neighbor. Note that we use only a few levels of segmentation {with about 1-8 segments per shard}. Inside each of the leaves, we build an HNSW graph which is known to give higher recall\cite{annbenchmarks}.


\subsubsection{Approximate Principal Direction Hyperplane Segmenters(APD)}
Approximate Principal Hyperplane Segmenters, are motivated by Approximate Principal Directions\cite{apdTrees} and Spectral Clustering\cite{trevisan2013lecture}. Since we would like to minimize the number of queries being routed to multiple segments, we propose using a spectral clustering instead of random hyperplanes. To speed up the process, we combine the principles of APD Trees\cite{apdTrees} and Spectral Clustering\cite{trevisan2013lecture}.

Let the dataset be denoted by $\mathcal{D}$ of dimensions $n \times d$. Let $A_{n\times n}$ denote the adjacency matrix of a similarity graph, $G$ constructed on $\mathcal{D}$. Let $D$ be the degree matrix of $A$ such that $D_{ii} = \sum_j A_ij$, and $C = D^{-1/2}AD^{-1/2}$. It is well-known that the largest eigenvalue of $C$ is 1, and the second-largest eigenvalue and the corresponding eigenvector approximate the sparsest cut \cite{trevisan2013lecture}. However, for large datasets, it is difficult to compute to the matrices $A$ and $C$ since they are of the order $O(n \times n)$. Along with these restrictions, we also have the added requirement of having a ``queryable" hyperplane.

To make this method work in practice, we assume that $A = \mathcal{DD}^T$ and that $\mathcal{D}$ is almost regular. The second-largest eigenvector of $A$ can be found using the second largest left singular vector of $\mathcal{D}$. Also, since $\mathcal{D} = U\Sigma V^T$ where $U$ and $V$ are the left and right singular vectors, we approximate the right singular vectors, as $U = \mathcal{D}.V$. Thus, we use $h$ which is the second-largest right singular vectors of $\mathcal{D}$, and let $U = \mathcal{D}.h$. 

Since we are using hyperplane based methods, this method also has the drawback of near points being across the splitting hyperplane. Thus, in this method as well, we employ the same methods of spill, insertion, and querying as described in Section~\ref{sec:RH}. Note that the theoretical guarantees from Section~\ref{sec:RH} are also applicable to the APD Segmenters. This bound may be loose since APD Segmenters make use of a data-dependent partitioning technique which can further improve the performance in practice.

\section{Offline Framework}
\label{sec:framework}
In this section, we describe the various components of LANNS. We propose pre-learning our learnable segmenters and feeding them as input to the indexing algorithm. The indexing algorithm stores the index on HDFS which can be fed into the querying algorithm, or can be exported to an online serving system (see Section~\ref{sec:online}).

\subsection{Learning a Segmenter}
\begin{figure}[ht]
    \centering
    \includegraphics[width=\linewidth]{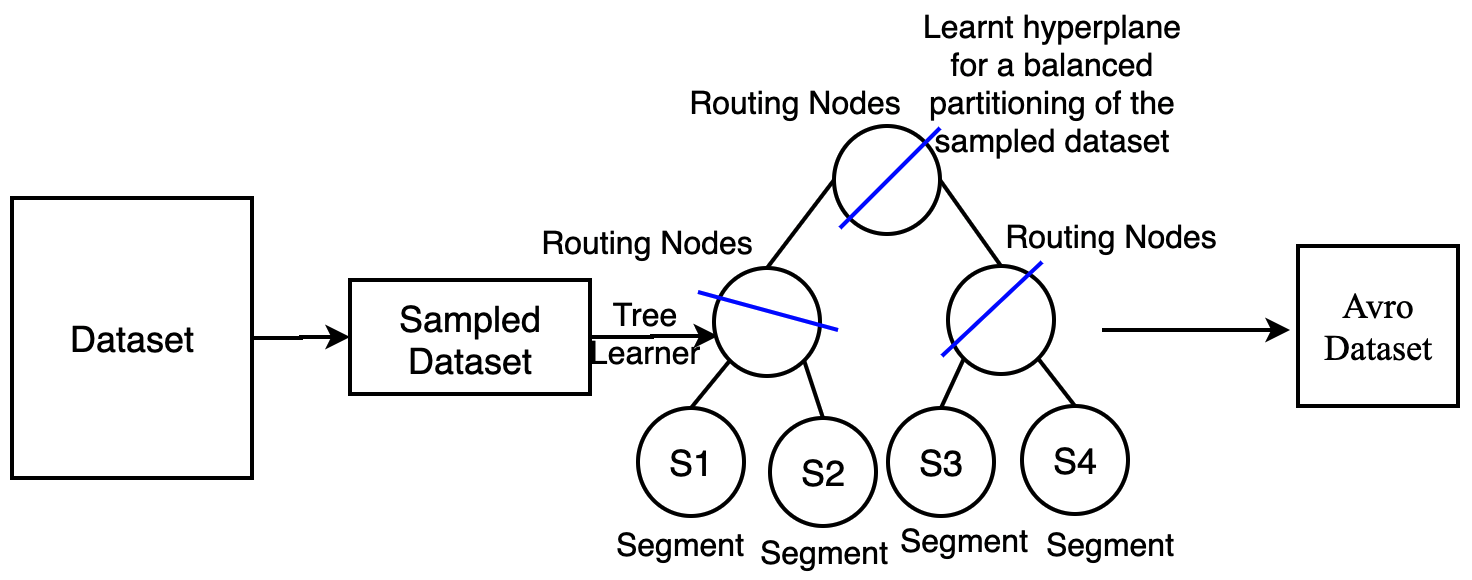}
    \caption{{Learning a Segmenter}}
    \label{fig:segmentation}
\end{figure}

Since the data distribution in our shards is uniform, we propose to pre-learn a segmenter and employ the same segmenter across all shards. This has a two-fold advantage-- (i) avoiding unnecessary computations to learn a segmenter for each shard on the fly. (ii) storing multiple segmenters for each shard in the offline system, since the segmenter is shared, only one copy of it is stored in the offline framework. In Figure~\ref{fig:segmentation}, we demonstrate our segmenter learning framework. Given the input dataset, we subsample the dataset uniformly at random. This sampled dataset, say, $\mathcal{D}$, is fed to the segmenter learning algorithm, which is one of Random Hyperplane Segmenter (RH) or Approximate Principal Direction Hyperplane Segmenter (APD). These techniques learn a tree of separating hyperplanes. At each internal node of this tree, a hyperplane is generated using RH or APD, which is used to further split the dataset into two partitions. For the APD Segmenter, we achieve a higher level of parallelism as the Spark Machine Learning library\cite{mllib} has an out of the box implementation available for Singular Value Decomposition. Once this tree of hyperplanes is learnt, we store the tree consisting of the hyperplane, the split points, and the left and right boundaries for each of the internal nodes. This learnt segmenter is fed to the ingestion algorithm described in the next section.

\subsection{Indexing}
\begin{figure}[ht]
    \centering
    \includegraphics[width=\linewidth]{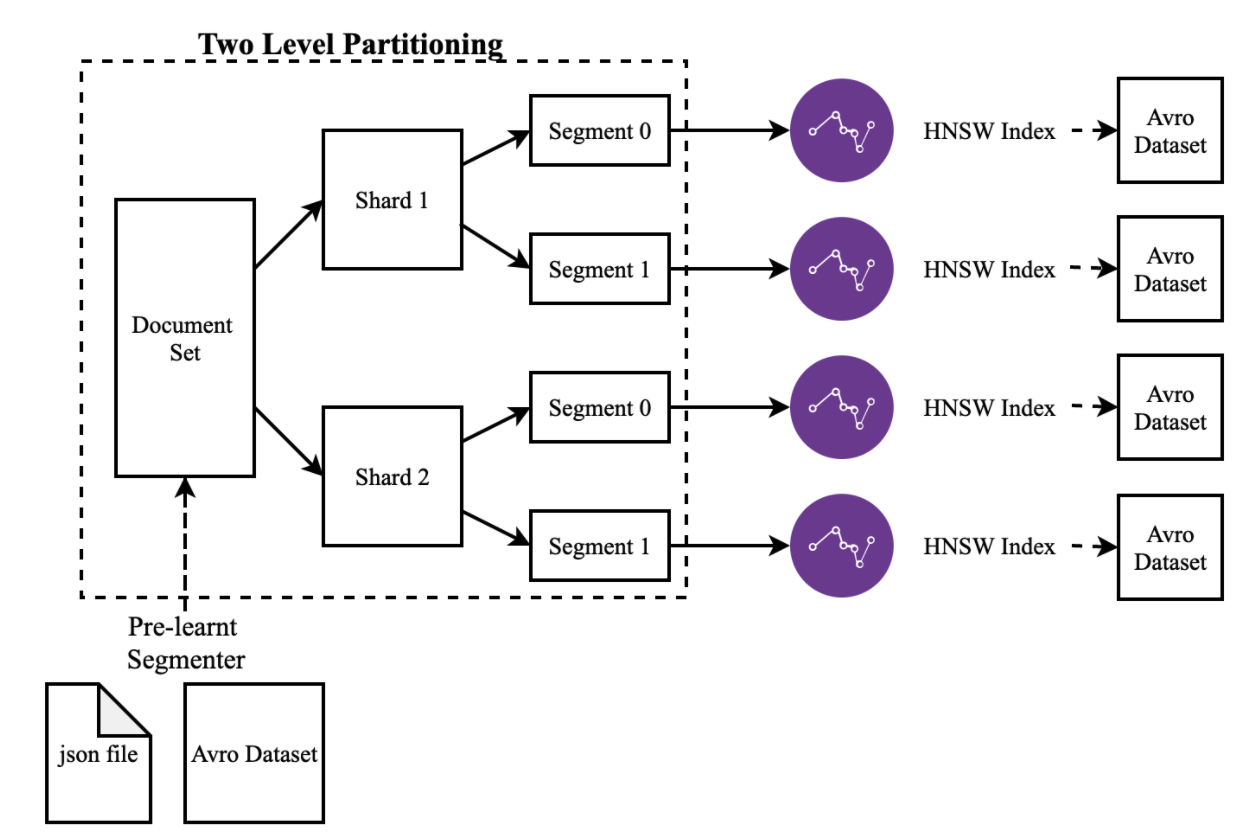}
    \caption{Indexing a multi-million sized dataset}
    \label{fig:indexing}
\end{figure}
In Figure~\ref{fig:indexing}, we show the process of scaling indexing for web-scale datasets. Along with the input dataset, we optionally input a pre-learnt segmenter. This segmenter is loaded within each of the Spark executors and is used to generate the two-level partitioning. This pre-learnt segmenter is shared across the shards. Each document is tagged with a shard Id and one or more segment Ids. The partition tagged dataset is then repartitioned based on these segment and shard Ids. One particular (shard, segment) pair is loaded in each executor and an HNSW Index is built on this subset. The HNSW Index is built inside the executor and hence, all these HNSW indexing can happen in parallel. The serialized index inside each executor is stored in the HDFS from the executor itself and the associated metadata and segmenter information is coupled with the index and written from the driver. 

\subsection{Querying}
\begin{figure*}[ht]
    \centering
    \includegraphics[width=0.9\linewidth]{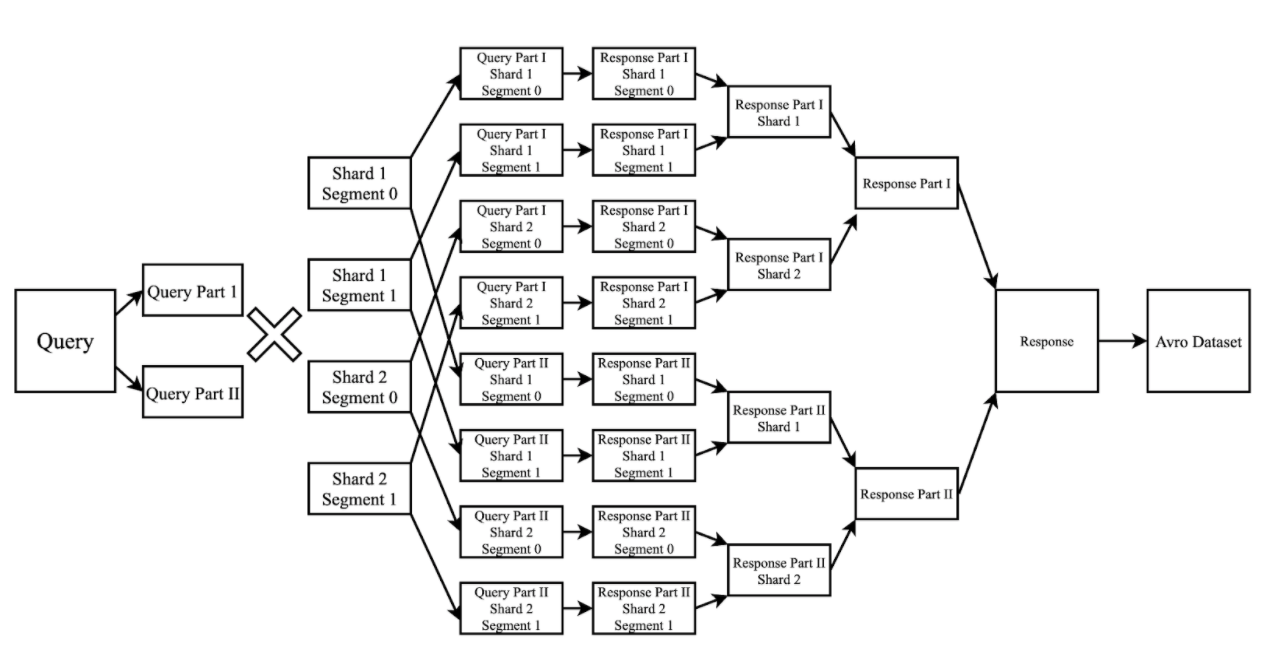}
    \caption{Querying with large query sets on multi-million datasets}
    \label{fig:querying}
\end{figure*}

For our offline use cases, it is of utmost importance to scale not only to big datasets but also to big query sets. To scale our query process, we make use of partitioned query sets. We demonstrate our process of querying with our two-level partitioned index in Figure~\ref{fig:querying}. We first take a large dataset and repartition these into smaller query partitions. These partitions are written to the HDFS. We also read the metadata of the index and prepare a `\texttt{SearchExecutorContext}` which informs each executor of which segment of which shard, and which query partition to load within in. This SearchExecutorContext is sent to the executors, the respective HNSW Indices and query partitions are loaded inside the executor. Partial search occurs inside each of these executors. We then employ a two-level merging as follows-- in the first level, partial results are returned to the driver along with the shard and segment Ids they are coming from. These partial results are repartitioned on the basis of the queryId and the shard Id to perform a segment level merging to obtain shard results. These shard results are then repartitioned again based on the query Ids for the final level of our two-level merging. This is analogous to how merging would occur in an online system. The segment results would first get merged within the server node containing the shard. These merged shard level responses are further merged from the searcher master/broker node.

\subsubsection{Preventing Time-Out Errors}
Spark occasionally suffers from time-out errors which could prove to be catastrophic in some large-scale systems. Consider a scenario where you have 100 \textit{(query, shard, segment)}-partitions, and only 8 executors. After performing partitioned searching, before the segment-level merging stage is triggered, some executors die and become unreachable. In these scenarios, the results become unavailable and the search for some partitions is restarted. While waiting for these recomputed results, some other executors may die, and so on. This leads to cascading failures which may cause catastrophic damages for time-sensitive applications. This is what we refer to as ``time-out`` errors. In order to prevent such scenarios from happening, we write partial results to a temporary path on the HDFS. After each phase of searching and merging, the results are written to a temporary path on the HDFS and are loaded from the temporary path for further processing. As soon as our two-level merging finishes, this temporary directory is cleaned. This works well since Spark ensures that for write operations, as soon as an executor finishes processing its task, instead of waiting for other executors to finish its task, it can write to the HDFS. This is in contrast to the repartitioning stages where the executor keeps the results and waits for all tasks of the stage to finish executing. 

\subsubsection{Per shard TopK}
Some recommender systems require searching for a very large number of nearest neighbors, of the order of 1000s, with further post-processing to prune candidates.
Sending the same "k" or "topK" to each shard can prove to be wasteful since each shard would then return topK responses. These topK responses would use up network I/O bandwidth and also increase the merge cost at the searcher or broker. In order to avoid such cases, we employ a "perShardTopK", which uses the Normal Approximation Interval\cite{confidenceInterval} to reduce the number of nearest neighbors fetched from each randomly partitioned shard. Let $S$ be the number of shards, and $p$ be the confidence (or topK.confidence), and $s' = \frac{1}{S}$, and
\begin{equation}
    cI = s' + f(p) * \sqrt{\frac{s' (1-s')}{topK}}
\end{equation}
then, 
\begin{equation}
    perShardTopK = min(topK, \lceil cI*topK \rceil)
\end{equation}

where $f(p)$ is the $(1- p/2)$ quantile of the standard normal distribution (i.e., the probit).

Note that since hyperplane based segmenters may query only a few segments, it is undesirable to apply the concept of a "per segment topK". Employing a per segment topK could lead to fewer than topK results as the final output. Thus, we do not optimize the topK for segments, instead we propagate the shard level perShardTopK to the associated segments.

\subsection{Brute Force Search}
\label{sec:bfs}
\label{sec:brute}
\begin{figure}[htb]
    \centering
    \includegraphics[width=\linewidth]{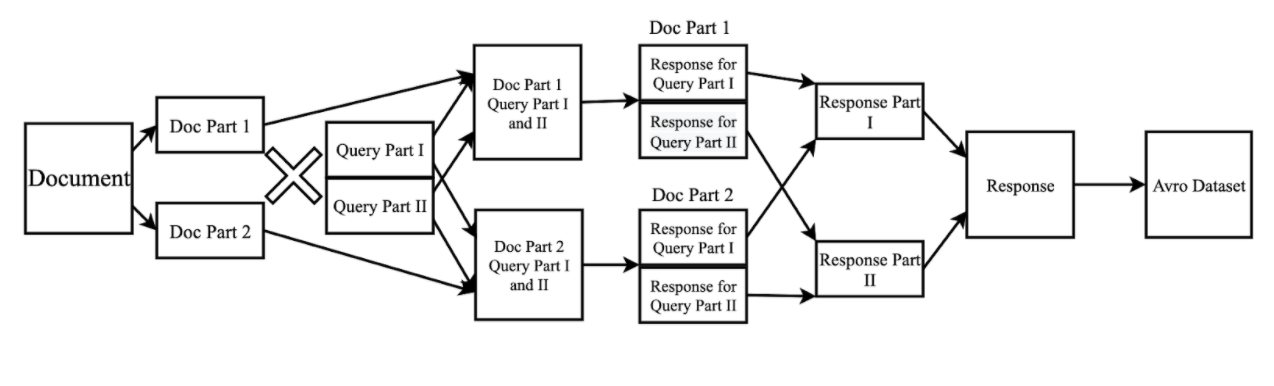}
    \caption{Brute Force Search on Large Datasets}
    \label{fig:brute}
\end{figure}

For benchmarking of LinkedIn specific datasets, we also implement brute force search on Spark to scale for large datasets. In Figure~\ref{fig:brute}, we show the process of scaling brute force search. We assume that the query set will be of reasonably small size and, first partition the dataset based on the number of executors available. Each subset of the dataset is loaded in the executor along with the whole query set. Partial results from each of these are computed and stored in the HDFS. From the driver, we once again load these partial results and repartition based on the query Id and merge results within executors. The merged results are then written to the HDFS for further recall computations. 

\section{Experiments}
\label{sec:expt}
\subsection{Open Source Evaluation}
For our evaluations on open-source data, we use two datasets-- (i) \texttt{SIFT1M}, the \texttt{SIFT1M} dataset with the indexing dataset of $1M$ records and query set of $10k$ records. Each of these have dimension, $d = 128$, and (ii) \texttt{GIST1M}, the \texttt{GIST1M}  dataset with the indexing dataset of $1M$ records and query set of $1k$ records. Each of these have dimension, $d = 960$. For both these datasets, we consider $topK=100$ nearest neighbors with the distance function to be Euclidean Distance. For both, \texttt{SIFT1M} and \texttt{GIST1M}, we compare our performance with HNSW algorithm. We build $(n,m)$-partitioned indices using Random Segmenters (RS), Random Hyperplane Segmenters (RH), and APD Segmenters (APD). For the \texttt{SIFT1M} dataset, we experiment with $(1,8)$-partitioned and $(2,4)$-partitioned indices.  For the \texttt{GIST1M} dataset, we limit ourselves to $(1,8)$-partitioned indices. For all experiments, we set $\alpha=0.15$, i.e., we route about $30\%$ queries to both partitions at any level. We also set the $topK.confidence = 0.95$ to limit the number of results obtained from each shard. For all experiments, the building times, query times, and recall are averaged over 5 runs.

\subsubsection{Results}

\begin{table}[ht]
\caption{Recall for the \texttt{SIFT1M} dataset. R@k refers to the Recall at topK = k. Method suffix $(n,m)$ refers to $(n,m)$-partitioning.}
\label{tab:siftRecall}

\begin{tabular}{|l@{}|r@{}|r|r|r|r|r@{}|}
\hline
\textbf{Method}   & {\textbf{R@1}} & {\textbf{R@5}} & {\textbf{R@10}} & {\textbf{R@15}} & {\textbf{R@50}} & {\textbf{R@100}} \\ \hline
\textbf{HNSW}     & \textbf{0.9912}                            & \textbf{0.9969}                            & \textbf{0.9977}                             & \textbf{0.998}                              & \textbf{0.9982}                             & \textbf{0.9981}                              \\ \hline
\textbf{RS(1,8)}  & 0.979                             & 0.9862                            & 0.9865                             & 0.9867                             & 0.987                              & 0.987                               \\ \hline
\textbf{RH(1,8)}  & 0.841                             & 0.818                             & 0.804                              & 0.798                              & 0.776                              & 0.762                               \\ \hline
\textbf{APD(1,8)} & 0.9772                            & 0.977                             & 0.975                              & 0.973                              & 0.9666                             & 0.9616                              \\ \hline
\textbf{RS(2,4)}  & \textbf{0.989}                             & \textbf{0.9944}                            & \textbf{0.995}                              & \textbf{0.995}                              & \textbf{0.996 }                             & \textbf{0.996}                               \\ \hline
\textbf{RH(2,4)}  & 0.9169                            & 0.9132                            & 0.9068                             & 0.9033                             & 0.8922                             & 0.885                               \\ \hline
\textbf{APD(2,4)} & \textbf{0.9898}                            & \textbf{0.9948}                            & \textbf{0.9944}                             & \textbf{0.9939}                             & \textbf{0.9926}                             & \textbf{0.9908}                              \\ \hline
\end{tabular}
\end{table}

\begin{table}[ht]
\caption{Build times for \texttt{SIFT1M} dataset with varying number of executors. Time is in minutes for total $1M$ data points.}
\label{tab:siftBT}

\begin{tabular}{|r|r|r|r|r|}
\hline
{\textbf{Executors}} & \textbf{HNSW}           & {\textbf{RS}} & {\textbf{RH}} & {\textbf{APD}} \\ \hline
\textbf{2}                            & \multicolumn{1}{r|}{40} & 8.2                              & 8.1                              & 8.4                               \\ \hline
\textbf{4}                            & -                       & 6.6                              & 6.8                              & 6.3                               \\ \hline
\textbf{8}                            & -                       & \textbf{4.3}                              & \textbf{4.4 }                             & \textbf{4.1}                               \\ \hline
\end{tabular}

\caption{Query times for \texttt{SIFT1M} dataset with varying number of executors. Time is in milliseconds per query for total $10k$ query points.}
\label{tab:siftQT}

\begin{tabular}{|@{}r|@{}r|r|r|r|r|r|r|}
\hline
& & \multicolumn{3}{l|}{\textbf{(1,8)-partitioning}} & \multicolumn{3}{l|}{\textbf{(2,4)-partitioning}} \\
{\textbf{Executors}} & \textbf{HNSW}            & {\textbf{RS}} & {\textbf{RH}} & {\textbf{APD}} & {\textbf{RS}} & {\textbf{RH}} & {\textbf{APD}} \\ \hline
\textbf{2}                            & \multicolumn{1}{r|}{50.4} & 58.8                              & 21                              & 16.8                               & 49.2                              & 46.8                              & 44.4                               \\ \hline
\textbf{4}                            & -                        & 46.2                              & 16.8                              & 12.6                               & 38.4                              & 25.8                              & 25.2                               \\ \hline
\textbf{8}                            & -                        & 25.8                              & 13.2                              & \textbf{10.2}                               & 33                              & \textbf{17.4 }                             & \textbf{17.4}                               \\ \hline
\end{tabular}
\end{table}

\begin{table}[ht]
\caption{Recall for the \texttt{GIST1M} dataset. R@k refers to the Recall at topK = k. Method suffix $(n,m)$ refers to $(n,m)$-partitioning.}
\label{tab:gistRecall}

\begin{tabular}{|l@{}|r|r|r|r|r|r@{}|}
\hline
\textbf{Method} & {\textbf{R@1}} & {\textbf{R@5}} & {\textbf{R@10}} & {\textbf{R@15}} & {\textbf{R@50}} & {\textbf{R@100}} \\ 
\hline
\textbf{HNSW}                        & 0.994                                    & 0.995                                    & 0.995                                     & 0.995                                     & 0.993                                     & 0.989                                      \\ \hline
\textbf{RS(1,8)}                          & \textbf{0.995}                                     & \textbf{0.998}                                    & \textbf{0.999}                                      & \textbf{0.999}                                     & \textbf{0.999}                                     & \textbf{0.999}                                      \\ \hline
\textbf{RH(1,8)}                         & 0.872                                    & 0.858                                    & 0.851                                     & 0.843                                     & 0.827                                     & 0.812                                      \\ \hline
\textbf{APD(1,8)}                         & 0.931                                    & 0.919                                    & 0.912                                     & 0.91                                      & 0.908                                     & 0.905                                      \\ \hline
\end{tabular}
\end{table}

\begin{table}[ht]
\caption{Build times for \texttt{GIST1M} dataset with varying number of executors. Time is in minutes for total $1M$ data points.}
\label{tab:gistBT}
\begin{tabular}{|r|r|r|r|r|}
\hline
{\textbf{Executors}} & \textbf{HNSW}            & {\textbf{RP}} & {\textbf{RH}} & {\textbf{APD}} \\ \hline
2                                                  & {577} & 132                              & 128                              & 140                               \\ \hline
4                                                  & -                        & 96                               & 108                              & 106                               \\ \hline
8                                                  & -                        & \textbf{48}                               & \textbf{54 }                              & \textbf{52}                                \\ \hline
\end{tabular}

\caption{Query times for \texttt{GIST1M} dataset with varying number of executors. Time is in milliseconds per query for total $1k$ query points.}
\label{tab:gistQT}

\begin{tabular}{|r|r|r|r|r|}
\hline
{\textbf{Executors}} & \textbf{HNSW}            & {\textbf{RP}} & {\textbf{RH}} & {\textbf{APD}} \\ \hline
2                                                  & {336} & 330                              & 156                              & 144                               \\ \hline
4                                                  & -                        & 222                              & 132                              & 108                               \\ \hline
8                                                  & -                        & 132                              & 96                              & \textbf{66}                               \\ \hline
\end{tabular}
\end{table}

In Tables~\ref{tab:siftRecall},\ref{tab:siftBT} and \ref{tab:siftQT}, we present the Recall, Build Times and Query Times comparisons with HNSW algorithm for the \texttt{SIFT1M} dataset. For Build and Query Times, we vary the number of executors for our segmented indices. With the random segmenter, RS, we observe comparable recall with $5\times$ speedup in build time and comparable query time with 2 executors, and $10\times$ speedup and a $2\times$ speedup query time with 8 executors. 
With the RH segmenter, we see a significant drop in the recall, $~15\%$ for $(1,8)$-partitioning. We observe speedups of $5x$ on the build time and $2.5\times$ on the query time with 2 executors, and $10\times$ on the build time and $4\times$ on the query time with 8 executors. With $(2,4)$-partitioning, we observe a $~8\%$ drop in recall with speedups of $5x$ on the build time and comparable query time with 2 executors, and $10\times$ on the build time and $3\times$ on the query time with 8 executors. 
For the APD segmenters, we observe $~2\%$ loss in recall with a $(1,8)$-partitioning with a $~5\times$ speedup in build time and $3\times$ speedup in query time with 2 executors, and $10\times$ speedup in build time and $5\times$ speedup in query time with 8 executors. With $(2,4)$-partitioning, we observe a $~1\%$ drop in recall with speedups of $5\times$ on the build time and comparable query time with 2 executors, and $10\times$ on the build time and $3\times$ on the query time with 8 executors. 
Build times do not change across $(1,8)$-partitioning and $(2,4)$-partitioning and across segmenters. This is because we pre-learn the segmenters and feed them to the ingestion setup. The RS doesn't require any pre-learning, RH segmenter takes $2.1$ minutes and $1.8$ minutes for $(1,8)$-partitioning and $(2,4)$-partitioning respectively on a subsample of 250k data points, and APD segmenter takes $3$ minutes and $2.6$ minutes for $(1,8)$-partitioning and $(2,4)$-partitioning respectively on a subsample of 250k data points. For all segmenter learning on \texttt{SIFT1M}, we use $10$ executors.

In Tables~\ref{tab:gistRecall},\ref{tab:gistBT} and \ref{tab:gistQT}, we present the Recall, Build Times and Query Times comparisons with HNSW algorithm for the \texttt{GIST1M} dataset. For Build and Query Times, we vary the number of executors for our segmented indices. With the random segmenter, RS, we observe comparable recall with $4.5\times$ speedup in build time with 2 executors and comparable query time and $11.5\times$ speedup in build time with 8 executors and a $2.5\times$ speedup in query time. 
With the RH segmenter, we see a significant drop in the recall, $~15\%$ for $(1,8)$-partitioning. We observe speedups of $5\times$ on the build time and $2\times$ on the query time with 2 executors, and $10\times$ on the build time and $3\times$ on the query time with 8 executors. 
For the APD segmenters, we observe $~7\%$ loss in recall with a $~5\times$ speedup in build time and $2\times$ speedup in query time with 2 executors, and $10\times$ speedup in build time and $5\times$ speedup in query time with 8 executors. 
The RS doesn't require any pre-learning, RH segmenter takes $6.3$ minutes on a subsample of 250k data points, and APD segmenter takes $18$ minutes on a subsample of 250k data points. For all segmenter learning on \texttt{GIST1M}, we use $30$ executors.

\subsection{Real-World Datasets}
We use four large-scale datasets for real-world use cases: 
\begin{enumerate}
    \item Groups Search: {\texttt{Groups}} is a dataset of $\sim$ 2.7M groups on LinkedIn. Each group is embedded in a 256-dimensional space. We evaluate the offline performance on a set of 10k queries.
    \item People Search: (\texttt{People}) is a database of 180M users of the LinkedIn platform. Each record is an embedding in 50 dimensions. This use case leverages our framework for people search on LinkedIn. We evaluate the offline performance on a set of 20k queries.
    \item People You May Know: (\texttt{PYMK}) is a database of 100M users of the LinkedIn platform. Each record is an embedding in 50 dimensions. We evaluate the offline recall performance on a subset of 1M queries, and offline query latency on 372M queries.
    \item Near-Duplicates: (\texttt{NearDupe}) consists of embeddings of multimedia images posted on the LinkedIn feed. The training set consists of 148k records in 2048 dimensions and the query set consists of 0.5M.
\end{enumerate}  

\texttt{Groups}, \texttt{People} and \texttt{NearDupe} use-cases were first tested using our offline platform and then onboarded to our online platform (See Section~\ref{sec:online}). \texttt{PYMK} use case, one of our biggest use-cases employs our offline platform for an in-production system. 

We first provide benchmarking results on the \texttt{Groups} dataset. For this, we use Python scripts to simulate our segmentation algorithm. We use the FAISS\cite{faiss} library to build an HNSW index inside each segment. We evaluate the following alternatives-- (i) Physical Spill: A data point lying close to the splitting plane is routed to both children or segments, (ii) Virtual Spill: A query which is close to the splitting plane is routed to both children or segments. The physical spill uses data side duplication and uses a higher memory footprint as compared to the virtual spill. However, the Queries per Second (QPS) in case of a physical spill is slightly higher than the virtual spill since the query is routed to only one segment in case of a physical spill. The results with both types of spill are presented in Table~\ref{tab:groups-spill}. For both of these, we see that the recall values are comparable with only a slight difference in the QPS values. For virtual spills, since queries are being routed to multiple segments, the number of unique queries that can be served is lower, this leads to a slight degradation in the QPS. 

For our use cases, since LANNS indices are used in production systems, it is undesirable to have a higher memory footprint. Employing physical spills for LANNS indices for large datasets such as \texttt{PYMK} would increase the memory footprint by about 30\% (~30M data points), which could increase the number of server nodes required. Thus, for our implementation, we use virtual spill even though it suffers from a slight drop in QPS as compared to physical spill. For the remaining results, we use virtual spill with $\alpha=0.15$.

\begin{table}[ht]
\caption{Recall on Groups dataset for a multi-segmented Index using APD Segmentation. R@k refers to the recall for k-nearest neighbors.}
\label{tab:groups-spill}
\begin{tabular}{|l|l|l|l|l|l|}
\hline
                           &                         & \multicolumn{2}{l|}{\textbf{Physical Spill}}          & \multicolumn{2}{l|}{\textbf{Virtual Spill}} \\ \hline
\textbf{\textbf{Segments}} & \textbf{\textbf{Spill}} & \textbf{\textbf{R@15}} & \textbf{\textbf{QPS}} & \textbf{R@15}   & \textbf{QPS}       \\ \hline
\textbf{1}                 & \textbf{0}\%            & \textbf{0.9458}               & \textbf{863.29}       & \textbf{0.9458}                 & \textbf{863.29}             \\ \hline
4                          & 10\%                    & 0.8400                        & 2619.02               & 0.8526                 & 2186.93            \\ \hline
4                          & 20\%                    & 0.8861                        & 2432.23               & 0.8853                 & 2010.44            \\ \hline
\textbf{4}                 & \textbf{30\%}           & \textbf{0.9268}               & \textbf{2392.42}      & \textbf{0.9272}        & \textbf{1984.21}   \\ \hline
8                          & 10\%                    & 0.7901                        & 2816.11               & 0.7866                 & 2852.21            \\ \hline
8                          & 20\%                    & 0.8510                        & 2774.32               & 0.8525                 & 2643.21            \\ \hline
\textbf{8}                 & \textbf{30\%}           & \textbf{0.9105}               & \textbf{2710.24}      & \textbf{0.9112}        & \textbf{2573}    \\ \hline
16                         & 10\%                    & 0.7359                        & 2993.32               & 0.7362                 & 3240.06            \\ \hline
16                         & 20\%                    & 0.8078                        & 2878.29               & 0.812                  & 3072.43            \\ \hline
\textbf{16}                & \textbf{30\%}           & \textbf{0.8836}               & \textbf{2797.42}      & \textbf{0.892}         & \textbf{2985.34}   \\ \hline
\end{tabular}
\end{table}

In Table~\ref{tab:real-times}, we present our building and querying times\footnote{Note that these times are inclusive of the times required for requesting a cluster and assigning executors.}\footnote{The building and querying time presented here are averaged over 5 runs.} for our real-world datasets. For the People and the PYMK use cases, owing to the large size of these datasets, it is infeasible to compare with the HNSW algorithm. For the NearDupe use cases, we essentially use the HNSW index with distributed querying. Therefore, we provide the results on our datasets with the parameters reflecting the optimal trade-off for our in-production services. We also present our recall evaluations in Table~\ref{tab:real-recall}. For each of the datasets, we obtain a high recall of over $~95\%$. All of these evaluations are performed on query sets of reasonable sizes. For large datasets, we employ an in-house Spark implementation of brute-force search as described in Section~\ref{sec:brute}. 

\begin{table}[ht]
\caption{Build and Query Times for Real-World datsets. S refers to number of shards, and dim refers to dimensions.}
\label{tab:real-times}
\begin{tabular}{|l|l|l|l|l|l|l|}
\hline
{ \textbf{}}        & \multicolumn{1}{l|}{{ \textbf{}}}       & & \multicolumn{2}{l|}{{ \textbf{Indexing}}}               & \multicolumn{2}{l|}{{ \textbf{Querying}}}               \\ \hline
{ \textbf{Dataset}} & \multicolumn{1}{l|}{{ \textbf{S}}} & \textbf{dim}& { \textbf{Size}} & { \textbf{Time}} & { \textbf{Size}} & { \textbf{Time}} \\ \hline
{ PYMK}             & { 20}     & 50                                       & { 100M}     & { 8h}            & {370M}     & {10h}        \\ \hline
{ People}           & { 32}     & 50                                       & { 180M}     & { 8h40m}        & {20k}      & {10m}           \\ \hline
{ NearDupe}         & { 1}      & 2048                                      & { 148k}   & { 1h20m}        & {500k}   & {5m}              \\ \hline
{ Groups}           & { 1}      & 256                                      & { 2.7M}    & { 2h13m}        & {20k}     & {7m}            \\ \hline
\end{tabular}
\end{table}

\begin{table}[ht]
\caption{Recall for Real-World datasets. S refers to number of shards, dim refers to dimensions, and R@K refers to the Recall at topK = K.}
\label{tab:real-recall}
\begin{tabular}{|l|l|l|l@{}|l@{}|l|l|}
\hline
\textbf{Dataset} & \textbf{S} & \textbf{dim}& \textbf{Index Size} & \textbf{Query Size} & \textbf{K} & \textbf{R@K} \\
\hline
\texttt{People}   & 32  & 50   & 180M   & 20k      & 50   & 97\%   \\
\hline
\texttt{PYMK}     & 20  & 50   & 100M    & 1M     & 100 & 95\%   \\
\hline
\texttt{NearDupe} & 1  & 2048    & 148k   & 0.5M   & 100  & 97\%   \\
\hline
\texttt{Groups} & 1 & 256 & 2.7M & 20k & 100 & 97\% \\
\hline
\end{tabular}
\end{table}

\section{Online Serving}
\label{sec:online}

\begin{figure}[ht]
    \centering
    \includegraphics[width=\linewidth]{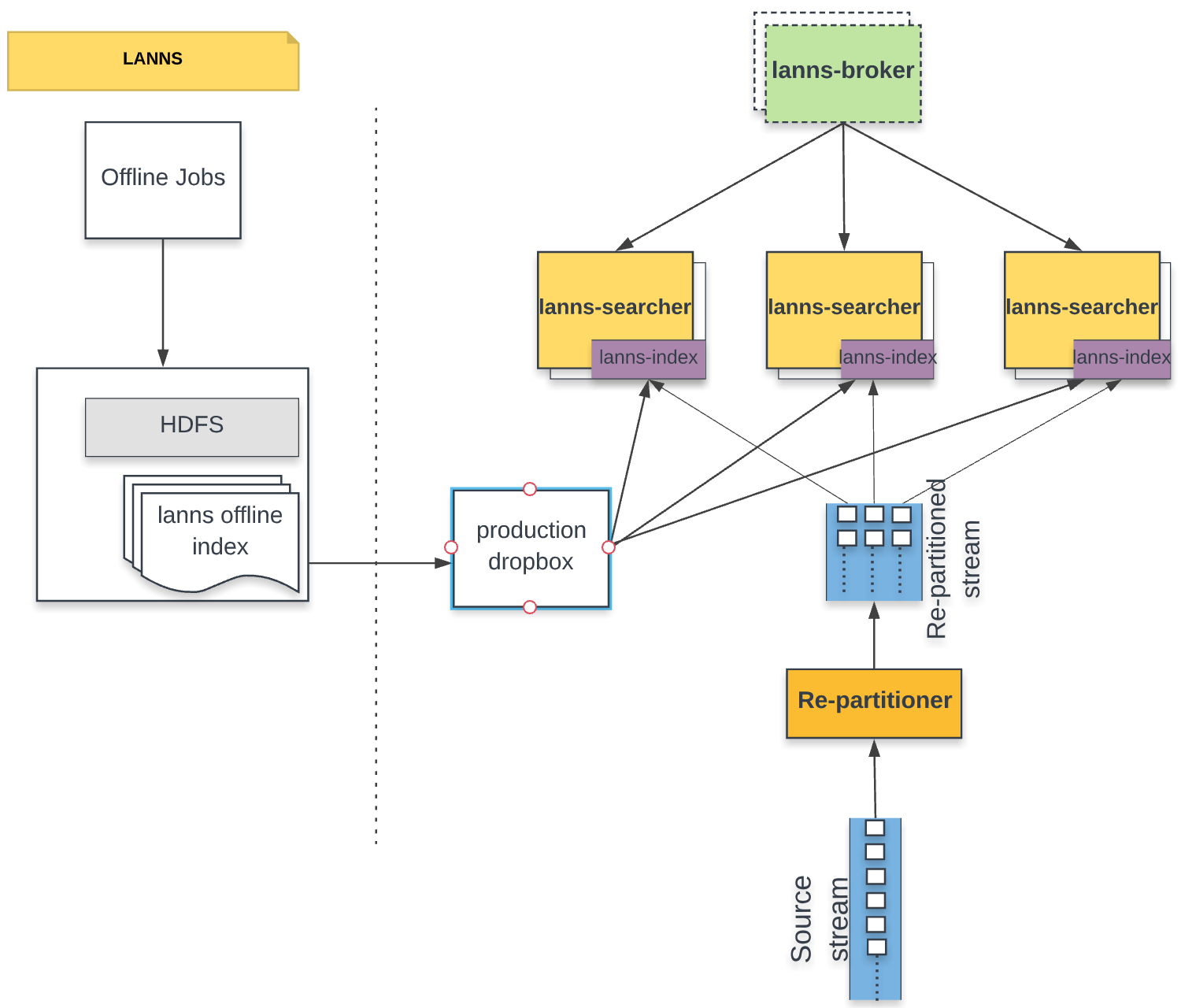}
    \caption{Online Service Architecture}
    \label{fig:online}
\end{figure}

Throughout the paper, we discuss the offline implementations and offline evaluations of our proposed method for scaling HNSW index builds. In this section, we will discuss our online service architecture, as shown in Figure~\ref{fig:online}. To enable nearest neighbor search capability in an online environment, we first build the index offline on our Spark cluster and then ship the serialized HNSW index (as Avro datasets) to online searcher nodes. The serialized index consists of the graph index, the actual embeddings (vectors) and additional metadata (like the segmenter, distance function used during index build, etc). The searcher nodes, when starting up, deserialize the index to native Java data-structures optimized for online serving using the persisted metadata with minimal additional configuration. This ensures that the platform doesn't allow accidental differences in the algorithm configuration between offline index build and online serving. The majority of storage needed in the online node comes from the vector representations of the entities in the index, the index itself is quite small. Fast lookup access to the embeddings for a document is critical for low latency online serving as most of the search time is spent on doing <query, document> distance comparisons. The difference in the online architecture is that each shard is hosted on a different machine. The first stage of the two-step merging, i.e., the shard level merging, happens at the machine where the shard is hosted (called a "searcher"), and the final merge happens at the broker or the client. The broker is also responsible for calculating and passing the $perShardTopK$ to each shard. We also built additional constructs to support use cases like hosting different indices in the same searcher to enable online A/B tests between different modeling techniques for creating the embedding representations of the documents in the index. For one of our large use cases with ~180M documents and embeddings of dimension 128, we benchmarked the online searcher to achieve a ~2.5K QPS at a p99 latency of ~20ms.

Our in-production systems make use of both, online as well as offline methods. One of our in-production use cases perform nearest neighbor search offline, using Spark, at some fixed cadence and send the processed results to online services for further processing. Three in-production use cases are hosted online with each shard hosted on a separate machine (or searcher). 

\textit{Offline v/s Online Serving} -- For applications with very large datasets and fixed query sets (for example, connections of members of any social network), we suggest offline processing and sending these processed results to online services. For very large datasets, the number of shards could be high as well. This would require several dedicated host machines, whereas, in offline mode, the processing is done only once and a cluster could be shared with various other applications. However, this cannot be done for applications where the query set is dynamic and the results are required instantly. For such cases, the online service is the only option.

\section{Conclusion}
\label{sec:conclusion}
In this work, we propose LANNS, an end-to-end platform for Approximate Nearest Neighbor Search. We enable scaling HNSW to web-scale datasets through a two-level partitioning scheme using the Spark framework. We demonstrate the excellent empirical performance on LinkedIn's production use-cases, and through extensive offline evaluations on various datasets. we demonstrate our scalable and highly competitive performance. We also briefly discuss the design choices and the trade-off between using an offline pipeline or an online, deployable service.

As future work, our approach of using segments can be explored for other purposes as well. For example, for context-based searches, we can build a segment per context and perform search in one or a few segments based on the contexts selected at query time.


\begin{acks}
We thank Liang Zhang for helping us get this project off ground. We also thank all the use case owners, Zhoutong Fu, Mohit Wadhwa, Nagaraj Kota, Sumit Srivastava and Yafei Wang for helping us make this project a success. 
\end{acks}

\bibliographystyle{ACM-Reference-Format}
\bibliography{acmart}

\appendix

\end{document}